# Effect of Electron Energy Distribution Function on Power Deposition and Plasma Density in an Inductively Coupled Discharge at Very Low Pressures

*Badri Ramamurthi and Demetre J. Economou*[1]

Plasma Processing Laboratory, Department of Chemical Engineering
University of Houston, Houston, TX 77204-4004

and

*Igor D. Kaganovich*[2]

Plasma Physics Laboratory, Princeton University, Princeton, NJ 08543

October 7, 2002

---

[1]e-mail: economou@uh.edu, [2]e-mail: ikaganov@pppl.gov

**Abstract**

A self-consistent 1-D model was developed to study the effect of the electron energy distribution function (EEDF) on power deposition and plasma density profiles in a planar inductively coupled plasma (ICP) in the non-local regime (pressure ≤ 10 mTorr). The model consisted of three modules: (1) an electron energy distribution function (EEDF) module to compute the non-Maxwellian EEDF, (2) a non-local electron kinetics module to predict the non-local electron conductivity, RF current, electric field and power deposition profiles in the non-uniform plasma, and (3) a heavy species transport module to solve for the ion density and velocity profiles as well as the metastable density. Results using the non-Maxwellian EEDF model were compared with predictions using a Maxwellian EEDF, under otherwise identical conditions. The RF electric field, current, and power deposition profiles were different, especially at 1mTorr, for which the electron effective mean free path was larger than the skin depth. The plasma density predicted by the Maxwellian EEDF was up to 93% larger for the conditions examined. Thus, the non-Maxwellian EEDF must be accounted for in modeling ICPs at very low pressures.

**1.0 INTRODUCTION**

Inductively coupled plasma (ICP) sources are used extensively for etching and deposition of thin films in microelectronics manufacturing. Such sources can produce a high-density, uniform plasma in a low pressure gas without the need for external magnetic fields [1,2,3,4,5].

At relatively high pressures (above ≈ 20 mTorr), electrons in an ICP discharge are heated by collisional dissipation of wave energy. Power deposition in lower pressure discharges, however, involves a collisionless electron heating mechanism [6,7]. It has been suggested that this is due to a "warm plasma" effect analogous to the anomalous skin effect in metals [4,8].

The anomalous skin effect in 1-D gas discharges has been studied theoretically for both semi-infinite and infinite systems [9,10,11]. Early experimental investigation of the skin effect was performed by Demirkhanov et al. [12]. The anomalous skin effect in 1-D *bounded* plasmas has also been studied both theoretically and experimentally [13,14,15,16]. An interesting effect associated with bounded plasmas is the possible resonance between the wave frequency and the motion of electrons bouncing between the walls. This can lead to enhanced (resonant) heating [15,17,18]. Most theoretical results reported thus far for a bounded plasma assume a uniform plasma density, where the electrostatic potential well is flat in the plasma and infinite at the wall (to simulate the existence of sheaths). In this square potential well, electrons are reflected back into the plasma only at the discharge walls. In a realistic non-uniform plasma, however, the electron turning points will depend on the electron total (kinetic plus potential) energy and the actual shape of the potential well, i.e., low total energy electrons bounce back at locations within the plasma. Although theoretical treatments of non-uniform slab plasmas have been reported [10,27], results related to such plasmas are lacking. A review of classical and recent works on the anomalous skin effect in plasmas was made in [19, 20].

At low pressures, when the electron mean free path is comparable to the discharge size, kinetic effects come into play, and the electron distribution function can be substantially non-Maxwellian [21].

Kinetic effects can be modeled using Monte Carlo or Particle-In-Cell (PIC) approaches, but these tend to be computationally intensive [22]. Besides, the direct use of conventional PIC-MCC for modeling of high-density ICP can be problematic due to the statistical noise in the charge and current density. In Ref.23 the traditional PIC approach was modified to reduced noise. This was based on consideration of correlations in electron motion with and without RF electric and magnetic fields. Additional reduction of the statistical noise was achieved using the condition of plasma quasineutrality.

Alternative "fast modeling" techniques [24,25,26] make use of analytical theory, and employ a number of simplifications, which can offer considerable improvement in computational time. This method has been successfully employed in the study of non-Maxwellian EEDFs in low-pressure RF capacitive discharges as well [24,25]. These approaches use the so-called quasi-linear theory [27,28], applied when the electron drift velocity is smaller than their thermal velocity, which is typical for low temperature discharges. However, nonlinear effects which arise due to interaction of electrons with transverse magnetic fields [29] near the boundary of the discharge can dampen electron heating, especially at the limit of electron-neutral collision frequency $\nu \to 0$ [30]. Such nonlinear effects were not treated in this article.

In a previous paper, a self-consistent model of non-local electron kinetics and heavy species transport in a 1-D slab (bounded) plasma was presented [31]. An argon discharge was studied incorporating both electron impact reactions and metastable chemistry. This model is used here to examine the effect of electron energy distribution function (EEDF) on power deposition profiles and plasma density. Special attention was paid to the effects of collisionless heating on the EEDF.

**2.0 MODEL DESCRIPTION**

A schematic of a 1-D parallel plate symmetric discharge (plate separation L) is shown in Fig. 1. Current sheets (not shown) on either side of the plasma, driven by a radio frequency (RF) source, generate a transverse RF field $E_y$ heating the plasma electrons. The RF field amplitude at the plasma edges is $E_0$; this value is set by the magnitude of the RF current and directly affects the total power deposited in the plasma. The RF field is attenuated by power transfer to the plasma electrons. Most of the power is deposited near the edge in what is called the "skin layer."

An electrostatic (space charge) field $E_{sc}(x)$ in the $x$-direction develops to confine electrons in the plasma and equalize the electron and ion current to the walls. The electron potential energy corresponding to this field is shown schematically in Fig. 2. Electrons with sufficiently low total ($x$-kinetic plus potential) energy will be reflected by this potential well. The reflection points $x_1^*$ and $x_2^*$ for an electron with total energy $\varepsilon$ are shown in Fig. 2. Thus, low energy electrons are confined near the discharge center, but higher energy electrons can reach further towards the walls. The sheath near the physical boundaries was not accounted for explicitly. Because of the high plasma density the sheath is only 100s of μm thick. Thus, the location of the sheath edge is essentially at the physical boundary, and the plasma approximation $n_i=n_e$ was applied to the whole domain. An infinite potential barrier was assumed for the sheath. Electrons with total energy higher than the potential at the sheath edge $\varphi_{sh}$ were assumed to reflect at the physical boundary, the underlying assumption being that the electron current to the wall was considered to be negligible.

Since non-local behavior is a warm plasma effect, kinetic treatment of electron transport is necessary. When electrons are warm enough to be transported out of the "skin layer" during an RF cycle, power is said to be deposited non-locally. In a sense, the current at a given location is influenced by the field at all other locations. In contrast, in the local case, the current at a given location only depends on the field at that particular point (Ohm's law). Non-locality is typically characterized by the parameter $l/\delta_0$, where $l = V_T/\sqrt{\omega^2 + \nu^2}$ is an "effective" electron mean free path, and $\delta_0$ is derived from the classical skin depth,

$$\delta_0 = \frac{c}{\omega_p}\left(1 + \frac{\nu^2}{\omega^2}\right)^{1/4}. \quad (1)$$

Here $V_T = \sqrt{2eT_e/m}$ is the most probable electron speed, $\omega_p$ is the electron plasma frequency ($\omega_p = \sqrt{e^2 n_e/m\varepsilon_0}$), $\omega$ is the RF frequency, $c$ is the speed of light in vacuum, $\nu$ (assumed constant) is the electron momentum-transfer collision frequency, $T_e$ is the electron temperature (in V), assuming a Maxwellian distribution function, and $m$ is the electron mass. By this definition, non-local behavior becomes significant when $l/\delta_0 \geq 1$.

For pressures typically smaller than 10 mTorr for argon, the electron energy distribution function (EEDF) can be non-Maxwellian [21]. Hence, for accurate calculation of power deposition and species density at low pressures, the EEDF needs to be computed. The following section describes a model for computing the EEDF for a low pressure argon plasma in which collisionless electron heating can be dominant [27,32].

**2.1 Electron Energy Distribution Function (EEDF) Module**

The Boltzmann equation for the EEDF $f$ (assuming a spatial dependence only in the $x$-direction) can be written as

$$\frac{\partial f}{\partial t} + v_x \frac{\partial f}{\partial x} - \frac{eE_{sc}(x)}{m}\frac{\partial f}{\partial v_x} - \frac{eE_y e^{i\omega t}}{m}\frac{\partial f}{\partial v_y} = S(f), \quad (2)$$

where, $E_{sc}(x)$ is the electrostatic field and $S(f)$ represents the sum of electron-atom (elastic and inelastic) and electron-electron collisions. For small deviations from the stationary EEDF $f_0$, one can write $f = f_0(x, v_x, v_y, v_z) + f_1(x, v_x, v_y, v_z, t)$, assuming that the relaxation time of the stationary EEDF $f_0$ is small compared to the RF period. Substituting for $f$ in Eq. (2) and integrating over the RF period, equations for $f_0$ and $f_1$ [27] are obtained,

$$v_x \frac{\partial f_0}{\partial x}\bigg|_{\varepsilon_x} = \left\langle \frac{eE_y(x,t)}{m}\frac{\partial f_1}{\partial v_y}\right\rangle + S(f_0), \quad (3)$$

and

$$-i\omega f_1 + v_x \frac{\partial f_1}{\partial x}\bigg|_{\varepsilon_x} = \frac{eE_y(x,t)v_y}{m}\frac{\partial f_0}{\partial \varepsilon} - \nu f_1, \quad (4)$$

where a harmonic dependence of the form $e^{-i\omega t}$ has been assumed for $f_1$. A new variable $\varepsilon_x$ (total energy in $x$-direction) is defined as $\varepsilon_x = mv_x^2/2e + \varphi(x)$. The brackets on the right hand side of Eq. (3) indicate averaging over the RF period.

Eq. (4) can be solved for $f_1$ by introducing a new variable $\theta$ [32] such that,

$$\theta = \Omega_b(\varepsilon_x)\int_{x_1^*}^{x}\frac{dx}{|v_x|}, \quad v_x > 0 \quad (5)$$

and

$$\theta = -\Omega_b(\varepsilon_x)\int_{x_1^*}^{x}\frac{dx}{|v_x|}, \quad v_x < 0, \quad (6)$$

where $\Omega_b(\varepsilon_x)$ is the bounce frequency of an electron with energy $\varepsilon_x$, given by

$$\Omega_b(\varepsilon_x) = \frac{\pi}{\int_{x_1^*}^{x_2^*}\frac{dx}{|v_x|}}, \quad (7)$$

$x_1^*$ and $x_2^*$ being the turning points corresponding to energy $\varepsilon_x$. Applying the above transformation to Eq. (4) yields,

$$-i\omega f_1 + \Omega_b(\varepsilon_x)\frac{\partial f_1}{\partial \theta} = E_y(x,t)v_y\frac{\partial f_0}{\partial \varepsilon} - \nu f_1. \quad (8)$$

Introducing the Fourier transform

$$f_{1n} = \int_{-\pi}^{\pi} f_1 e^{in\theta}d\theta \quad (9)$$

and solving for $f_1$ in Eq. (8),

$$f_{1n} = \frac{E_{yn}v_y}{(\Omega_b(\varepsilon_x)ni - i\omega + \nu)}\frac{\partial f_0}{\partial \varepsilon}, \quad (10)$$

where

$$E_{yn}(\varepsilon_x) = \frac{\Omega_b(\varepsilon_x)}{\pi}\int_0^{\pi}\frac{E_y(x)\cos(n\theta(x))}{|v_x|}dx. \quad (11)$$

Eq. (3) can be further simplified utilizing the fact that the electron energy is approximately conserved for $\nu \ll \omega, \Omega_b$. Then, Eq. (3) can be averaged over the bounce time $T_b(\varepsilon_x)$ and the "perpendicular velocities" $v_y$ and $v_z$ to yield

$$\frac{\partial}{\partial \varepsilon}\left(D_\varepsilon(\varepsilon)\frac{\partial f_0}{\partial \varepsilon}\right) = \overline{S(f_0)} \quad (12)$$

where $D_\varepsilon(\varepsilon)$ is the energy diffusion coefficient given by [32]

$$D_\varepsilon(\varepsilon) = \frac{\pi}{8}\left(\frac{2e}{m}\right)^{3/2} \times \\ \sum_{n=-\infty}^{\infty}\int_0^{\varepsilon}d\varepsilon_x\left|E_{yn}(\varepsilon_x)\right|^2\frac{\nu(\varepsilon-\varepsilon_x)}{\Omega_b(\varepsilon_x)\left(\left[\Omega_b(\varepsilon_x)n-\omega\right]^2+\nu^2\right)} \quad (13)$$

In the limit of $\nu \gg \Omega_b$, which corresponds to the case of collisional heating, Eq. (13) can be shown [33] to reduce to

$$D_\varepsilon(\varepsilon) = \frac{1}{6}\left(\frac{2e}{m}\right)\int_{x_1^*(\varepsilon)}^{x_2^*(\varepsilon)}\left|E_y(x)\right|^2(\varepsilon-\varphi(x))^{3/2}\frac{\nu}{\nu^2+\omega^2}dx,$$

where, $x_1^*$ and $x_2^*$ are the turning points for an electron with energy $\varepsilon$.

The right hand side of Eq. (12) denotes the space- and bounce time-average of electron-atom and electron-electron collisions [33]. The collisions considered in this model were elastic electron-atom collisions, inelastic electron-atom collisions (ground-state ionization and excitation, metastable ionization) and electron-electron collisions. Consequently, $\overline{S(f_0)}$ was written as

$$\overline{S(f_0)} = \overline{S_{el}(f_0)} + \overline{S_{iz}(f_0)} + \overline{S_{ex}(f_0)} + \overline{S_{mi}(f_0)} + \overline{S_{ee}(f_0)} \quad (15)$$

where subscripts 'el', 'iz', 'ex', 'mi' and 'ee' on the right hand side stand for elastic, ground-state ionization, excitation, metastable ionization and electron-electron collisions respectively.

Each of the above spatially averaged terms can be written as [33]

$$\overline{S_{el}(f_0)} = \frac{d}{d\varepsilon}\left(\overline{V}(\varepsilon)f_0\right) \quad (16)$$

$$\overline{S_{iz}(f_0)} = \sqrt{\frac{2e}{m}} \times \\ \left(2\overline{u\nu_{iz}^*(u)}f_0(\varepsilon) - 2\overline{(u+\varepsilon_{iz})\nu_{iz}^*(u+\varepsilon_{iz})}f_0(\varepsilon+\varepsilon_{iz})\right) \quad (17)$$

$$\overline{S_{ex}(f_0)} = \sqrt{\frac{2e}{m}} \times$$
$$\left( \overline{u\nu_{ex}^*(u)} f_0(\varepsilon) - \overline{(u+\varepsilon_{ex})\nu_{ex}^*(u+\varepsilon_{ex})} f_0(\varepsilon+\varepsilon_{ex}) \right)$$
(18)

$$\overline{S_{mi}(f_0)} = \sqrt{\frac{2e}{m}} \times$$
$$\left( \overline{u\nu_{mi}^*(u)} f_0(\varepsilon) - \overline{(u+\varepsilon_{mi})\nu_{mi}^*(u+\varepsilon_{mi})} f_0(\varepsilon+\varepsilon_{mi}) \right)$$
(19)

$$\overline{S_{ee}(f_0)} = \Gamma \frac{d}{d\varepsilon} \left( \bar{H}(\varepsilon) f_0 + \frac{2}{3} \bar{G}(\varepsilon) \frac{df_0}{d\varepsilon} \right) \quad (20)$$

where, the coefficients $\bar{V}(\varepsilon), \bar{H}(\varepsilon)$ and $\bar{G}(\varepsilon)$ are given by [33]

$$\bar{V}(\varepsilon) = \kappa \sqrt{\frac{2e}{m}} \int_{x_1^*(\varepsilon)}^{x_2^*(\varepsilon)} u^{3/2} \nu_m(u) dx \quad (21)$$

$$\bar{H}(\varepsilon) = \int_{x_1^*(\varepsilon)}^{x_2^*(\varepsilon)} dx' \int_0^u \sqrt{u'} f_0(u') du' \quad (22)$$

$$\bar{G}(\varepsilon) = \int_{x_1^*(\varepsilon)}^{x_2^*(\varepsilon)} dx' \left[ \int_0^u u'^{3/2} f_0(u') du' + u^{3/2} \int_u^\infty f_0(u') du' \right] \quad (23)$$

The factor "2" was used in the first term in parenthesis on the right hand side of Eq. (17) to model the loss of electrons to the walls, i.e., for every electron produced due to an ionization event, one electron is lost to the walls. Variable $u$ in Eqs. (21)-(23) represents the kinetic energy of electrons, $u = \varepsilon - \varphi(x)$, and should not be confused with the ion fluid velocity introduced later.

The pre-factor $\Gamma$ in Eq. (20) depends on the Coulomb logarithm $\Lambda$

$$\Gamma = \frac{2e}{m} \frac{e^2}{8\pi\varepsilon_0} \ln(\Lambda) \quad (24)$$

Substituting Eqs. (15)-(23) in Eq. (12), one obtains the final form of the equation for the stationary EEDF $f_0(\varepsilon)$

$$-\frac{\partial}{\partial \varepsilon} \left[ D_\varepsilon(\varepsilon) + \frac{2\Gamma}{3} \bar{G}(\varepsilon) \right] \frac{\partial f_0(\varepsilon)}{\partial \varepsilon} -$$
$$\frac{\partial}{\partial \varepsilon} \left[ \bar{V}(\varepsilon) + \Gamma \bar{H}(\varepsilon) \right] f_0(\varepsilon) = \overline{S_{iz}(f_0)} + \overline{S_{ex}(f_0)} + \overline{S_{mi}(f_0)} \quad (25)$$

Boundary conditions for Eq. (25) were specified for large energies, assuming that both $f_0$ and $\partial f_0 / \partial \varepsilon$ are "small" (note that both cannot be zero as the integration of the discretized form of Eq. (25) would not proceed). The exact values for $f_0$ and $\partial f_0 / \partial \varepsilon$ are not important as the distribution function was normalized such that

$$\int_0^\infty \sqrt{\varepsilon} f_0(\varepsilon) d\varepsilon = 1 \quad (26)$$

Eq. (25) was solved as an initial value problem, starting with an initial condition at $\varepsilon = 75$ V and marching backwards to $\varepsilon = 0.01$ V, using a fourth order Runge-Kutta scheme.

### 2.2 Non-local Electron Kinetics (NLEK) Module

This module solves for the RF electric field, current and power deposition profiles in a non-uniform plasma. In the non-local regime, the current at any point in the plasma depends on the electric field at all other points. Maxwell's equations can be reduced to a single scalar equation for the transverse electric field $E_y$ [31],

$$\frac{d^2 E_y}{dx^2} = i \left( \frac{\omega_{p0}}{c} \right)^2 \frac{\omega}{2\sqrt{2e/m}} \times$$
$$\left( \int_0^x G(x,x') E_y(x') dx' + \int_x^L G(x',x) E_y(x') dx' \right) \quad (27)$$

where $\omega_{p0} = (e^2 n_{e0} / m \varepsilon_0)^{1/2}$ is the electron plasma frequency evaluated using the peak electron density $n_{e0}$. The boundary conditions are $E_y(0) = E_y(L) = E_0$. The time-average power deposition profile can be computed as

$$P(x) = \frac{1}{2} \mathrm{Re} \left( J_y(x) E_y^*(x) \right) \quad (28)$$

where $E_y^*(x)$ is the complex conjugate of $E_y(x)$, and Re is the real part of the quantity in parenthesis. The conductivity kernel $G(x,x')$ depends on the profile of the potential well confining electrons in the plasma [31] which has to be computed self-consistently as part of the simulation. The expression for the RF current $J_y(x)$ was also given in [31].

### 2.3 Heavy Species Transport (HST) Module

The heavy species transport module solves for the ion density and velocity profiles as well as the metastable species density. Since the extremely thin sheaths were not included in the simulation, the quasi-neutrality constraint was imposed, and the location of the plasma-sheath boundary was taken to be the wall. Since the drift-diffusion approximation for ions is questionable at pressures below about 10 mTorr, a momentum equation was solved to compute the Ar$^+$ velocity. The metastable Ar* species density is quite uniform at pressures below about 10 mTorr. Hence, a spatially average (0-D) model was used to determine Ar*. Due to symmetry, only half the domain ($0 \leq x \leq L/2$) was considered.

The HST module uses the mass continuity equation for Ar$^+$ and the ion momentum equation written in terms of ion velocity,

$$\frac{\partial u_+}{\partial t} + u_+ \frac{\partial u_+}{\partial x} = -\frac{eT_{eff}(x)}{m_+} \frac{\partial (\ln n_+)}{\partial x} - \frac{eT_+}{m_+} \frac{\partial (\ln n_+)}{\partial x} - \nu_+(u_+)u_+ - \frac{(R_i + R_{mi} + R_{mp})}{n_+} u_+, \quad (29)$$

Here, $u_+$, $n_+$, and $m_+$ are ion velocity, density, and mass respectively. $T_{eff}(x)$ is the "screening temperature." Reaction rates $R_i$, $R_{mi}$, and $R_{mp}$ correspond to ground state ionization, metastable ionization, and metastable pooling, respectively (Table I). The second term on the RHS of Eq. (29) can be neglected compared with the first term since $T_+ \ll T_e$. For the collisional drag (third) term on the RHS, a constant mean-free path was employed, whereby the ion-neutral collision frequency as a function of ion velocity $\nu_+(u_+)$ was written as $\nu_+(u_+) = \nu_{+0}|u_+|/u_{+,th}$; $\nu_{+0}$ is a reference collision frequency (Table II) at which the ion drift velocity equals the ion thermal velocity $u_{+,th} = \sqrt{eT_+/m_+}$. The fourth term on the RHS represents a "drag" in the sense that the ions that are being produced by ionization have negligible drift velocities, and have to be brought up to the local drift velocity. The ionization rate $R_i$ in Eq. (29) was calculated through the EEDF

$$R_i(x) = N_0 \sqrt{\frac{2e}{m}} \int_{\varphi(x)}^{\infty} \sigma_I(u)(\varepsilon - \varphi(x)) f_0(\varepsilon) d\varepsilon, \quad (30)$$

where $N_0$ is the gas density, and $\sigma_I(u)$ is the ionization cross-section as a function of electron kinetic energy $u = \varepsilon - \varphi(x)$. A corresponding expression was used for $R_{mi}$.

The boundary condition for ion velocity was set at the wall as

$$u_+ = -u_b = -\sqrt{eT_{eff}(0)/m_+}, x = 0 \quad (31)$$

where the Bohm velocity is $u_b$. Due to symmetry, there was no ion flux at the discharge center ($x=L/2$). (see Ref. 31 for more details on the HST module.)

### 2.4 Maxwellian EEDF Calculation

The model described in the previous sections can be used for the self-consistent calculation of RF discharge properties in the non-local regime including a *non-Maxwellian* EEDF. In order to compare this model with one employing a Maxwellian EEDF, the following modifications were made.

1) The non-local conductivity $G(x,x')$ was written for a (normalized) Maxwellian distribution $f_0(\varepsilon) = 2\pi/(\pi T_e)^{3/2} e^{-\varepsilon/T_e}$.

2) An electron energy equation (based on electron temperature) was added to the HST module

$$\frac{\partial}{\partial t}\left(\frac{3}{2} n_e T_e\right) = -\frac{\partial q_e}{\partial x} + P(x) - 3\frac{m}{m_n} n_e \nu (T_e - T_g) - \sum_j R_{je} \Delta H_{je} \quad (32)$$

where the electron energy flux $q_e = -K_e \partial T_e/\partial x + 5/2 \Gamma_e T_e$, and the electron mass flux $\Gamma_e = \Gamma_i = n_+ u_+$. $P(x)$ is the power density profile obtained from the non-local electron kinetics module (Eq. 28). The third term on the right hand side (RHS) of Eq. (32) is electron energy loss in elastic collisions; $m_n$ is the heavy species (Ar) mass, and $T_g$ is the neutral gas temperature. The last (summation) term accounts for energy loss (or gain) in inelastic collision $j$ with rate $R_{je}$ and energy exchange $\Delta H_{je}$ (Table I). Boundary conditions were:

$$q_e = \frac{5}{2} \Gamma_e T_e, x = 0$$
$$q_e = 0, x = L/2 \quad (33)$$

3) The reaction rates for electron-neutral collisions were obtained using expressions of the form of Eq. (30), $f_0(\varepsilon) = 2\pi/(\pi T_e)^{3/2} e^{-\varepsilon/T_e}$.

### 3.0 METHOD OF SOLUTION

The simulation consisted of three modules: an electron energy distribution function (EEDF) Module, a non-local electron kinetics (NLEK) module, and a heavy species transport (HST) module (Fig. 3). The NLEK module computed the non-local conductivity kernel $G(x,x')$ and solved for the RF electric field and current profiles. The RF electric field at the wall $E_0$ was adjusted to match the target total power. The RF field profiles were used in the EEDF module to compute the energy diffusion coefficient $D_\varepsilon(\varepsilon)$ (Eq. (13)) and solve for the distribution function $f_0(\varepsilon)$. The latter was used to compute the electron-impact reaction rate coefficients via Eq. (30) (e.g., ionization and excitation rates), and the effective electron temperature ($T_{eff}$) profile. The effective electron temperature was used in the ion momentum equation (Eq. 29), and also in the boundary condition for ion velocity at the wall (Eq. 31). The ionization and excitation rates were used as source terms in the continuity equations for ions, and metastables. The HST module provided, among other quantities, the ion (electron) density and electrostatic potential $\varphi$ as a function of position. These were fed back to the NLEK module to calculate a new RF field profile. The calculation was repeated until convergence to a self-consistent solution. When using a Maxwellian EEDF, iterations were performed between the NLEK and HST module only (the EEDF module was not used). In this case, the HST module included an electron energy balance (Eq. 32) to compute the electron temperature profile.

The *initial* electron density profile was assumed to be a sine function peaking at the center. The corresponding potential was computed assuming a uniform Maxwellian temperature of 2.5 V. Convergence was declared when the potential profile changed by less than 0.1% (in the $L_2$ norm), which typically took about 30 iterations around the modules. At convergence, the ion density profile predicted by the HST module and the electron density profile predicted by the EEDF module differed by less than 0.1% (in the $L_2$ norm). The computation time on a 933 MHz Intel Pentium 3 Windows NT was ~ 10 hrs for a run with non-Maxwellian EEDF, and ~ 1 hr for a run with Maxwellian EEDF.

## 4.0 RESULTS AND DISCUSSION

Base-case parameter values and constants used in the simulation are shown in Table II. Results in Figs. 4-8 are for a pressure of 1 mTorr and discharge frequency of 13.56 MHz. Under these conditions the electron collision frequency is small compared to the applied field frequency. Results in Figs. 9-13 are for a pressure of 10 mTorr and discharge frequency of 13.56 MHz. Under these conditions, the electron collision frequency is comparable to the applied field frequency. In each case, profiles calculated using the non-Maxwellian EEDF module (solid lines) are compared with profiles (dashed lines) obtained using the Maxwellian EEDF approximation under the same discharge conditions and for the same (integrated) *total* power. Values of power correspond to a plate cross sectional area of $64\pi$ cm$^2$.

### 4.1 Pressure=1 mTorr

Fig. 4a shows the EEDF as a function of total energy for non-Maxwellian (solid lines) and Maxwellian (dashed lines) cases. The non-Maxwellian EEDF has a higher fraction of electrons just beyond the ionization threshold, predicting a higher ionization rate. For a pressure of 1 mTorr, the electron collision frequency $\nu \sim 3 \times 10^6$ s$^{-1}$ and $\nu/\Omega_b \sim 0.1$. The energy diffusion coefficient $D_\varepsilon(\varepsilon)$ (Eq. (13)), exhibits a "knee" at ~ 1 V (Fig. 4b), indicating that the "temperature" of electrons with energies less than 1 V is lower than that of electrons with energies greater than 1 V. The "knee" in Fig. 4b arises due to a phenomenon called "bounce heating" or "resonant heating". For $\nu/\Omega_b \sim 0.1$, the energy diffusion coefficient in Eq. (13) can be approximated as

$$D_\varepsilon(\varepsilon) \approx \frac{\pi}{8}\left(\frac{2e}{m}\right)^{3/2} \sum_{n=1}^\infty \int_0^\varepsilon d\varepsilon_x |E_{yn}(\varepsilon_x)|^2 \times$$
$$(\varepsilon - \varepsilon_x)\frac{\delta[\Omega_b(\varepsilon_x)n - \omega]}{\Omega_b(\varepsilon_x)}, \tag{34}$$

where, $\delta(\Omega_b(\varepsilon_x)n - \omega)$ represents the Dirac-delta function. It can be seen from Eq. (34) that for energy $\varepsilon < \varepsilon_1$ (where $\varepsilon_1$ is obtained from $\Omega_b(\varepsilon_1) = \omega$), $D_\varepsilon(\varepsilon) \approx 0$. For $\varepsilon_1 < \varepsilon < \varepsilon_2$ (where $\Omega_b(\varepsilon_2) = \omega/2$), $D_\varepsilon(\varepsilon) \propto (\varepsilon - \varepsilon_1)$, i.e., the energy diffusion coefficient increases linearly with total energy. This behavior leads to the "knee" observed in Fig. 4b, and implies that electrons with energy $\sim \varepsilon_1$ (in this case is ~1 V) are in resonance with the field and are thus heated more efficiently. Higher order resonant modes ($n$ =2,3,4...) contribute less as the Fourier coefficients $E_{yn}$ decrease as $n$ increases. Resonant heating has been demonstrated for capacitively coupled plasmas [26], where the electron density is relatively smaller and so the resonance effect can be more pronounced. The dashed line in Fig. 4b shows the dependence on total energy of the e-e collision diffusivity term (given by $D_{ee}(\varepsilon) = 2/3\Gamma\bar{G}(\varepsilon)$ in Eq. (25)).

Fig. 5 shows the effective temperature profiles for the Maxwellian and non-Maxwellian EEDFs. For the Maxwellian case, the electron temperature is independent of power while for the non-Maxwellian case, significant differences are observed with power. The large difference between the temperatures at the edge and the center may be explained by examining Fig. 4a (solid lines). The EEDF shows that electrons with total energies less than ~1 V are not effectively heated. Electrons with such low energies are essentially trapped near the discharge center (where the heating field is weak) as they cannot overcome the electrostatic potential barrier. Hence, the effective temperature at the center is low. In contrast, electrons with relatively high energies can overcome the potential barrier and reach the edge where the field is strong, and the effective temperature at the periphery (and larger total energies) is high. Note that even for the highest plasma density in Fig. 4, the electron-electron mean free path is about 10 m for 1 eV electrons, much higher than the interelectrode gap. Therefore, the electron-electron and collisionless energy diffusion coefficients are comparable at very low energy, ~ 1 eV, see Fig. 4. As a result, low energy electrons form a Maxwellian distribution with very low temperature, ~ 1 eV [34]. Note that the part of the EEDF corresponding to such cold electrons is difficult to measure experimentally.

The effective temperature profile becomes less non-uniform as power is increased, because of higher electron density resulting in more "thermalization" of the distribution by e-e collisions. The discrepancy between the Maxwellian temperature and the effective temperature near the edge induces a difference in the effective electron mean free path, which leads to considerably different field and current density profiles as discussed below.

Fig. 6 shows the profiles of the normalized amplitude of the RF field. The field profile is monotonic for low power. However, for high power, the behavior becomes progressively non-monotonic due to increasing non-locality. Specifically, the skin depth decreases with power, and the more energetic electrons can escape from the skin layer during a RF cycle, resulting in non-local behavior and non-monotonic RF field profiles. The effect of non-locality is more pronounced for the non-Maxwellian EEDF, especially for higher powers for which the RF field at the discharge center is more than 50% of the value at the edge. This is a direct consequence of the higher effective temperature predicted by the non-Maxwellian EEDF near the edge compared to the Maxwellian case. Warmer electrons can reach further in the discharge core.

The corresponding power deposition profiles are shown in Fig. 7. The peak of power deposition in the Maxwellian case is seen to occur closer to the boundary, when compared to that of the non-Maxwellian case. This is because of the higher effective temperature of electrons in the skin layer for the non-Maxwellian case, which causes them to travel a greater distance during an RF cycle. Both cases exhibit negative power deposition near the discharge center. This can be explained by the phase difference between the current and the RF field; electrons can pick up energy from the field within the skin layer and lose energy back to the field outside the skin layer. Negative power deposition has been observed experimentally for low-pressure inductively coupled discharges [35].

The corresponding positive ion density profiles are shown in Fig. 8. The positive ion density is determined by two factors: (1) the effective electron temperature at the boundary, which controls the loss rate of ions to the wall (Eq. 31) and (2) the rate of ionization (ground-state and metastable). The latter depends on the tail of the EEDF beyond the ionization threshold of 15.76 V (ground state ionization dominates under these conditions). The ionization rate was found to be marginally higher for the non-Maxwellian EEDF. However, the effective temperature at the wall for the non-Maxwellian case (~ 6.5 V) is larger than the Maxwellian temperature of 4.4 V (Fig. 5), leading to larger losses for the non-Maxwellian EEDF. This results in lower density for the non-Maxwellian case. The differences in the peak densities are 32.4, 38.8, and 44.4%, respectively, for 50, 100, and 200 W.

### 4.2 Pressure=10 mTorr

For a pressure of 10 mTorr, the electron collision frequency is $\nu = 3 \times 10^7$ s$^{-1}$ which is comparable to the discharge frequency of $\omega = 8.52 \times 10^7$ s$^{-1}$. Fig. 9 shows the EEDF as a function of total energy for the non-Maxwellian (solid lines) and Maxwellian (dashed lines) cases. The collision frequency $\nu > \Omega_b$, where $\Omega_b$ is the electron bounce frequency. Consequently, in contrast to 1 mTorr, there is no resonant (bounce) heating of low energy electrons. Furthermore, since the electron density at 10 mTorr is sufficiently high, electron-electron collisions tend to thermalize the bulk of the distribution. This results in a bi-Maxwellian EEDF with bulk electrons having an effective temperature of ~3 V, and electrons at the edge having a temperature of ~3.5 V (Fig. 10). Fig. 10 also suggests that the distribution function is more Maxwellian-like at higher powers (higher electron density), leading to a less non-uniform effective temperature profile. This effect is more pronounced at higher pressures (compare to Fig. 5), again due to higher electron density. An important difference between Maxwellian and non-Maxwellian EEDFs in Fig. 9 is the depletion of high energy electrons in the non-Maxwellian case.

Fig. 11 shows the normalized profile of the amplitude of the RF field as a function of position for total power of 50, 100 and 200 W. Solid and dashed lines correspond to the non-Maxwellian and Maxwellian EEDFs, respectively. Interestingly, the Maxwellian case predicts more "non-locality" at 10 mTorr compared to 1 mTorr (Fig. 6), where the opposite was true. This is due to the considerably higher electron density (Fig. 13) predicted by the Maxwellian EEDF, leading to smaller skin depth and more non-locality.

The corresponding power deposition profiles are shown in Fig. 12. The profiles for the Maxwellian and non-Maxwellian cases are similar, with the power deposition reaching a maximum within the skin layer, and decaying towards the center of the discharge. The peak in power density for the Maxwellian case occurs ~ 0.4 cm from the wall, while the peak location is seen to vary for the non-Maxwellian EEDF. This is because the temperature of the Maxwellian EEDF does not change with power, but the effective temperature of the non-Maxwellian EEDF does vary with power (Fig. 10).

The corresponding positive ion density profiles are shown in Fig. 13. Significant differences are observed with the Maxwellian EEDF predicting higher densities. The deviation in peak density for 50, 100, and 200 W is 70.5, 88.2, and 92.8%, respectively. The reason for the discrepancy is twofold: (a) the Maxwellian EEDF predicts more ionization (tail extending to higher energies, Fig. 9), and (b) the effective electron temperature at the edge is lower for the Maxwellian EEDF (Fig. 10), leading to lower ion losses (Eq. 31).

### 4.3 Comparison with Experiments

Figures 14 and 15 show comparisons between simulated and experimental EEDFs for an asymmetric 2D inductively coupled Ar discharge [36]. The chamber ID was 19.8 cm and the inter-electrode spacing was 10.5 cm. The driving frequency of the coil current was 13.56 MHz. Similar experimental results were obtained in Ref. [37]. Because experimental data were taken in a reactor which is at least 2-D, while the model presented in this work is only 1-D, the EEDF module was tested separately, using the fact that the EEDF is determined mostly by the total power deposition. A uniform electron density was assumed (square potential well) corresponding to the peak density (at the discharge center), obtained experimentally [36]. Other parameters were the electron-neutral collision frequency and the RMS value of the RF field $E_0$ at the wall [36]. An RF field corresponding to the local approximation (Ohm's law) was used for this comparison. These values were used to compute the energy diffusion coefficient (see Eq. (13)) and then obtain the EEDF (Eq. (25)).

Fig. 14 shows the comparison between experimental (open symbols) and simulated (lines) EEDFs for a pressure of 1 mTorr. The experimental profiles are clearly non-Maxwellian and this fact is reflected in the simulated profiles. The agreement between theory and experiment is very good for 200 W where the experimental behavior is captured for almost the entire energy range. This is not the case for 12 W and 50 W where theory predicts a longer high-energy tail. However, the temperature of the tail agrees well with the experimental temperature. Similar trends are observed in Fig. 15, which is for 10 mTorr.

It should be once more noted that the model presented in this work is 1-D while the experiments were performed in a system which is at least 2-D. Hence, a self-consistent simulation of the experiment is not possible. For this reason, comparisons were made using the experimentally measured plasma density. The measured skin depths were also used for the RF field with the local approximation. The goal of the exercise was to test the theoretical EEDF model as a stand-alone module. The discrepancies between theory and experiment might well be due to the multidimensional nature of the experiment, which is not captured by the model.

## 5.0 CONCLUSIONS

A self-consistent 1-D model was developed to study the effect of electron energy distribution function on power absorption and plasma density profiles in a planar inductively coupled argon discharge in the non-local regime (pressures ≤ 10 mTorr). The model consisted of three modules: (1) an electron energy distribution function (EEDF) module to compute the non-Maxwellian EEDF, (2) a non-local electron kinetics module to predict the non-local electron conductivity, RF current, electric field and power deposition profiles in the non-uniform plasma, and (3) a heavy species transport module to solve for the ion density and velocity profiles as well as the metastable density. The self-consistent simulation predicted the RF electric field, power deposition, electron energy distribution function, and ion density profiles. Results using the non-Maxwellian EEDF were compared with those predicted by assuming a Maxwellian EEDF under otherwise identical conditions.

The self-consistently determined EEDFs for the non-Maxwellian and Maxwellian cases were quite different for both 1 and 10 mTorr. At a pressure of 1 mTorr, the non-Maxwellian EEDF showed a "resonant heating" mechanism, where electrons with energies greater than 1 V were heated more efficiently, compared to the Maxwellian case. This suggests that "resonant heating" can be important in determining the EEDF at very low pressures. At 10 mTorr, the non-Maxwellian EEDF assumed a bi-Maxwellian structure, with "temperatures" of ~3.5 V and ~3 V, respectively, for the low energy and high-energy parts of the distribution. A depletion of the high-energy tail was observed in the non-Maxwellian EEDF due to electron energy losses in inelastic collisions. The power deposition profile for the non-Maxwellian and Maxwellian cases were different for the pressures and powers investigated. This was attributed to different temperatures predicted by the two EEDF models. In both cases, negative power deposition was observed near the center of the discharge, due to non-local electron kinetics. The ion (electron) density predicted by the Maxwellian EEDF was up to 93% larger than that predicted by the non-Maxwellian EEDF. This was mainly due to a larger effective electron temperature at the wall predicted by the non-Maxwellian EEDF, leading to higher ion losses. Thus, accounting for the non-Maxwellian EEDF is important for modeling ICPs operating at very low pressures.

**ACKNOWLEDGEMENTS** BR and DJE are thankful to the National Science Foundation (CTS 0072854) for financial support. IDK acknowledges the Plasma Physics Laboratory of Princeton University for financial help through a "University Research Support Program."

**Table I**: Reactions used in the argon discharge simulation [38]. $\Delta H_j$ is electron energy loss (positive value) or gain (negative value) upon collision.

| No. | Process | Symbol | Reaction | $\Delta H_j$ (eV) |
|---|---|---|---|---|
| R1 | Ground state excitation | $R_{ex}$ | $Ar + e \rightarrow Ar^* + e$ | 11.6 |
| R2 | Ground state ionization | $R_i$ | $Ar + e \rightarrow Ar^+ + 2e$ | 15.8 |
| R3 | Step-wise ionization | $R_{si}$ | $Ar^* + e \rightarrow Ar^+ + 2e$ | 4.2 |
| R4 | Superelastic collisions | $R_{sc}$ | $Ar^* + e \rightarrow Ar + e$ | -11.6 |
| R5 | Metastable quenching | $R_{mq}$ | $Ar^* + e \rightarrow Ar^r + e$ | |
| R6 | Metastable pooling | $R_{mp}$ | $Ar^* + Ar^* \rightarrow Ar^+ + Ar + e$ | |
| R7 | Two-body quenching | $R_{2q}$ | $Ar^* + Ar \rightarrow 2Ar$ | |
| R8 | Three-body quenching | $R_{3q}$ | $Ar^* + 2Ar \rightarrow Ar_2 + Ar$ | |

**Table II**: Parameter values used in the simulation

| Parameter | Value |
|---|---|
| Plasma length, $L$ | 5 cm |
| Ion temperature, $T_i$ | 0.026 V |
| Plate area, $A$ | $64\pi$ cm$^2$ |
| Reference ion collision frequency, $\nu_{i0}$ (@ $3.2\ 10^{14}$ cm$^{-3}$) | $1.66\ 10^5$ s$^{-1}$ |
| Electron momentum-exchange collision frequency, $\upsilon$ (@ $3.2\ 10^{14}$ cm$^{-3}$) | $3\ 10^7$ s$^{-1}$ |
| Gas Temperature | 300 K |
| Excitation frequency | 13.56 MHz |

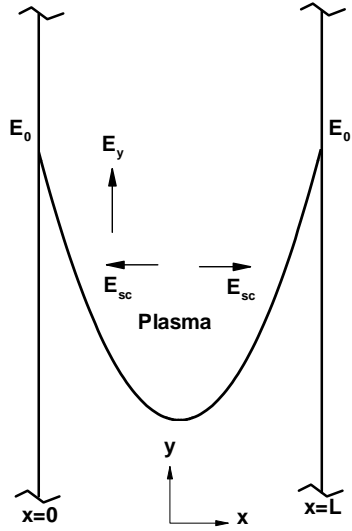

**Figure 1**: Schematic of a one-dimensional plasma slab of length *L* powered by a symmetric inductively coupled source. The RF current source (not shown) results in an RF field in the transverse direction, $E_y$. The value of the field at the edges, $E_0$, is determined by the desired power deposition in the plasma. A space charge field $E_{sc}$ develops in the x-direction to confine electrons.

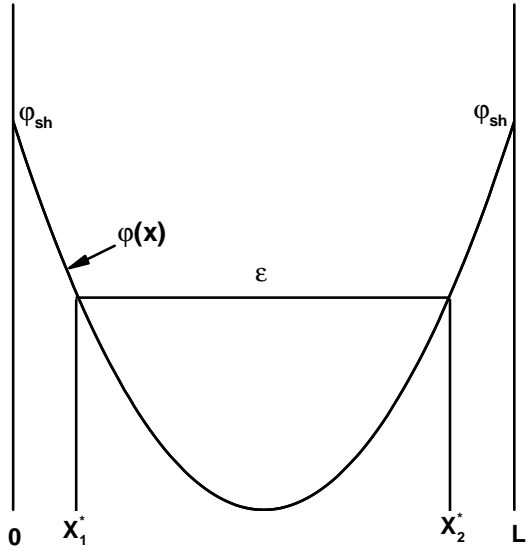

**Figure 2**: Schematic of electron potential energy profile $\varphi(x)$ in the plasma slab, due to the electrostatic field. An electron with total (*x*-kinetic plus potential) energy $\varepsilon$ will reflect back at points $x_1^*$ and $x_2^*$ (turning points).

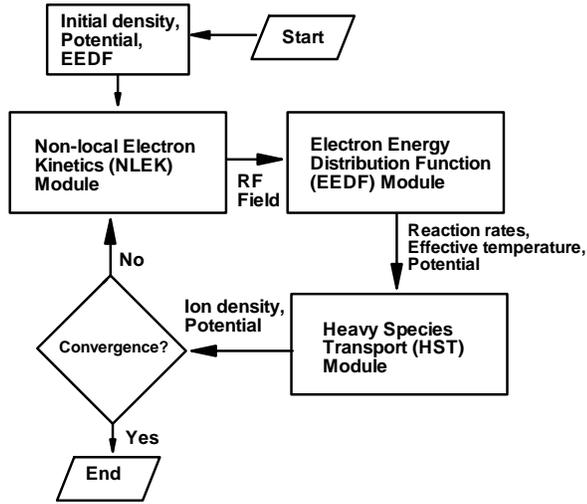

**Figure 3**: Flow diagram used for numerical simulation. Simulation cycled between the three modules until the potential and plasma density profiles converged.

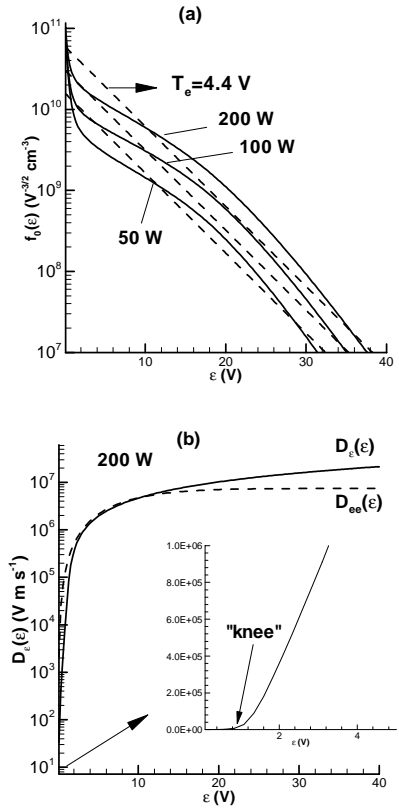

**Figure 4**: (a) Self-consistently predicted non-Maxwellian (solid lines) and Maxwellian (dashed lines) electron energy distribution function (EEDF) as a function of total energy for 1 mTorr. (b) Energy diffusion coefficient (Eq. 13) $D_\varepsilon(\varepsilon)$ (solid line) and energy diffusivity (see text) related to e-e collisions (dashed line) as a function of total energy for 1 mTorr. Inset shows an expanded scale for $D_\varepsilon(\varepsilon)$. Other conditions as in Table II.

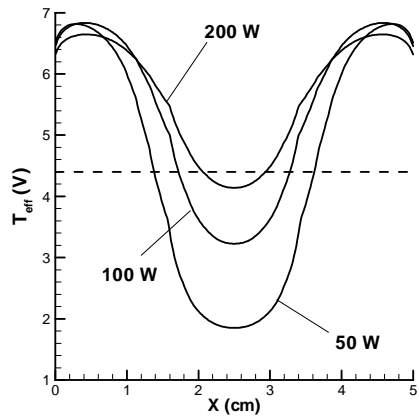

**Figure 5**: Effective temperature profiles for a non-Maxwellian EEDF (solid lines) and a Maxwellian EEDF (dashed lines) for 1 mTorr. Other conditions as in Table II.

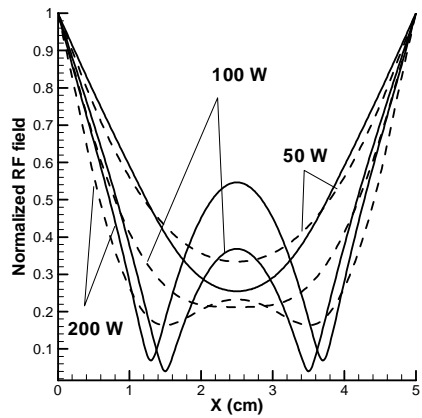

**Figure 6**: Normalized amplitude of the RF field for 1 mTorr. Results using non-Maxwellian EEDF (solid lines) are compared with results using Maxwellian EEDF (dashed lines), under otherwise identical conditions. Other conditions as in Table II.

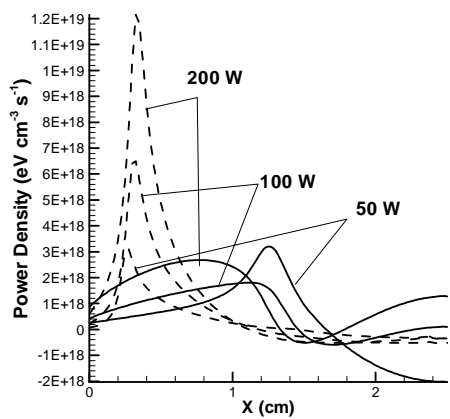

**Figure 7**: Power density profiles for 1 mTorr. Results using non-Maxwellian EEDF (solid lines) are compared with results using Maxwellian EEDF (dashed lines), under otherwise identical conditions. Other conditions as in Table II.

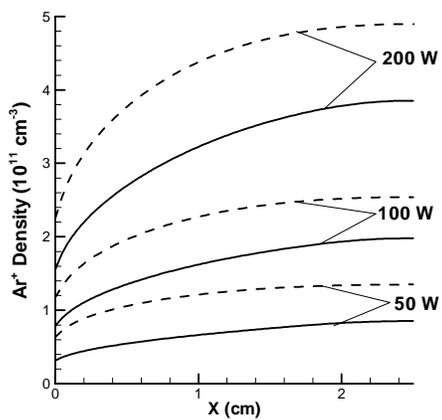

**Figure 8**: Variation of positive ion density for 1 mTorr. Results using non-Maxwellian EEDF (solid lines) are compared with results using Maxwellian EEDF (dashed lines), under otherwise identical conditions. Other conditions as in Table II.

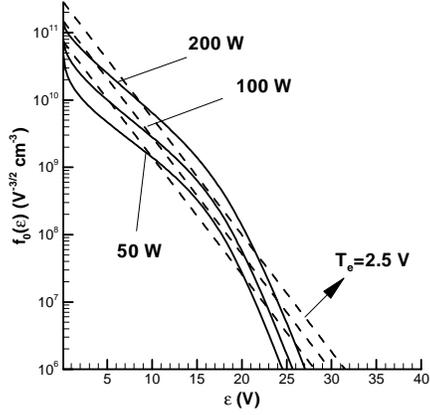

**Figure 9**: Self-consistently predicted non-Maxwellian (solid lines) and Maxwellian (dashed lines) electron energy distribution function (EEDF) as a function of total energy for 10 mTorr. Other conditions as in Table II.

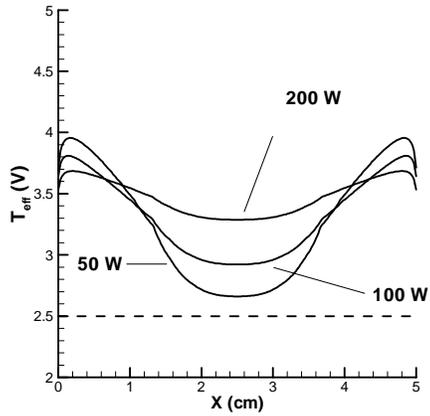

**Figure 10**: Effective temperature profiles are shown for non-Maxwellian (solid lines) and Maxwellian (dashed lines) EEDF for 10 mTorr. Other conditions as in Table II.

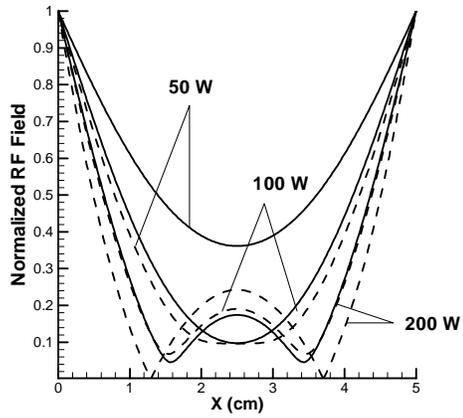

**Figure 11**: Normalized amplitude of the RF field for 10 mTorr. Results using non-Maxwellian EEDF (solid lines) are compared with results using Maxwellian EEDF (dashed lines), under otherwise identical conditions. Other conditions as in Table II.

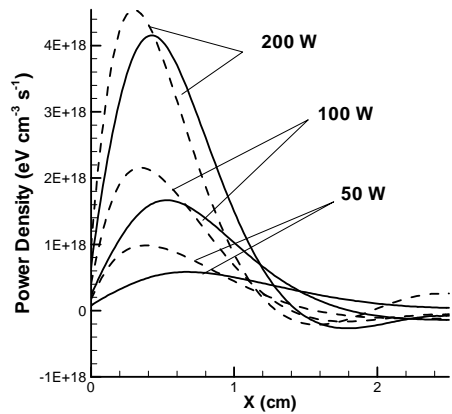

**Figure 12**: Power density profiles for 10 mTorr. Results using non-Maxwellian EEDF (solid lines) are compared with results using Maxwellian EEDF (dashed lines), under otherwise identical conditions. Other conditions as in Table II.

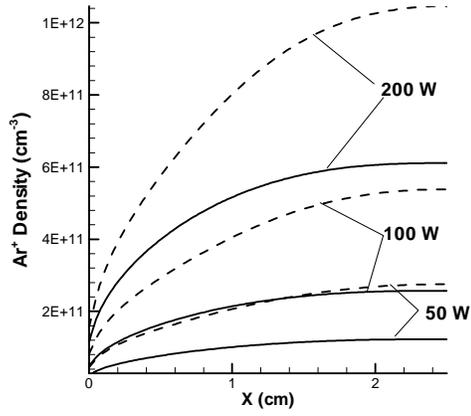

**Figure 13**: Positive ion density for 10 mTorr. Results using non-Maxwellian EEDF (solid lines) are compared with results using Maxwellian EEDF (dashed lines), under otherwise identical conditions. Other conditions as in Table II.

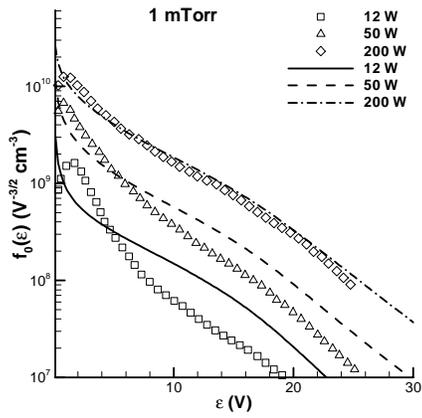

**Figure 14**: Comparison between simulated (lines) and experimental (symbols) EEDF for 1 mTorr. Data from ref. [36].

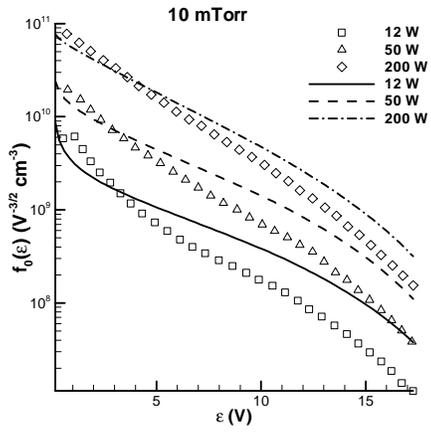

**Figure 15**: Comparison between simulated (lines) and experimental (symbols) EEDF for 10 mTorr. Data from ref. [36].